\documentclass[aps,twocolumn,groupedaddress,amssymb,amsfonts,showpacs]{revtex4}

\usepackage{color}
\usepackage[dvips]{graphics,graphicx}
\newcommand{\shortqph}[1]{}

\providecommand{\ignore}[1]{}

\def\Z{{\mathbb{Z}}}

\def\proof{{\em Proof:  }}

\def\openone{\leavevmode\hbox{\small1\kern-3.8pt\normalsize1}}
\def\RR{{\rm I\kern-.2emR}}
\def\tr{{\rm tr}\; }

\def\proof{{\em Proof:  }}

\def\openone{\leavevmode\hbox{\small1\kern-3.8pt\normalsize1}}
\def\RR{{\rm I\kern-.2emR}}
\def\tr{{\rm tr}\; }

\providecommand{\ignore}[1]{}

\newcommand{\ket}[1]{| #1 \rangle}
\newcommand{\bra}[1]{\langle #1 |}
\newcommand{\proj}[1]{\ket{#1}\! \bra{#1}}

\newcommand{\bitem}{\begin{itemize}}
\newcommand{\eitem}{\end{itemize}}
\newcommand{\benum}{\begin{enumerate}}
\newcommand{\eenum}{\end{enumerate}}
\newcommand{\beq}{\begin{equation}}
\newcommand{\eeq}{\end{equation}}
\newcommand{\beqa}{\begin{eqnarray}}
\newcommand{\eeqa}{\end{eqnarray}}
\newcommand{\bi}{{\bf i}}
\newcommand{\bj}{{\bf j}}
\newcommand{\bk}{{\bf k}}
\newcommand{\bl}{{\bf l}}

\newtheorem{definition}{Definition}

\newtheorem{proposition}{Proposition}

\newcommand{\bproof}{\begin{proof}}
\newcommand{\eproof}{\end{proof}}
\newcommand{\bprop}{\begin{proposition}}

\newcommand{\bdef}{\begin{definition}}

\begin{document}


\title{Symmetry and Topological Order}



\author{Zohar Nussinov$^1$ and Gerardo Ortiz$^{2}$}
\affiliation{$^1$Department of Physics, Washington University, St.
Louis, MO 63160, USA}
\affiliation{$^2$Department of Physics, Indiana University, Bloomington,
IN 47405, USA}

\date{\today}  

\begin{abstract}

\end{abstract}


\begin{abstract}
We prove sufficient conditions for Topological Quantum Order at both zero and finite temperatures. The crux of the proof hinges on the existence of low-dimensional Gauge-Like Symmetries (that notably extend and differ from standard local gauge symmetries) and their associated defects, thus providing a unifying framework based on a symmetry principle. These symmetries may be actual invariances of the system, or may emerge in the low-energy sector. Prominent examples of Topological Quantum Order display Gauge-Like Symmetries. New systems exhibiting such symmetries include Hamiltonians depicting orbital-dependent spin exchange and Jahn-Teller effects in transition metal orbital compounds, short-range frustrated Klein spin models, and p+ip superconducting arrays. We analyze the physical consequences of Gauge-Like Symmetries (including topological terms and charges), discuss associated braiding, and show the insufficiency of the energy spectrum, topological entanglement entropy, maximal string correlators, and fractionalization in establishing Topological Quantum Order. General symmetry considerations illustrate that not withstanding spectral gaps, thermal fluctuations may impose restrictions on certain suggested quantum computing schemes and lead to ``thermal fragility''. Our results allow us to go beyond standard topological field theories and engineer systems with Topological Quantum Order. 
\end{abstract}
\maketitle

The role of invariance (symmetry) principles in accounting for observed
regularities is well known \cite{wigner}.  Invariances become
particularly effective in quantum mechanics where the linear character
of Hilbert space enables us to  construct superpositions of states that
transform as irreducible representations of various symmetry groups. The
first discovered invariance principles were of a {\it geometric
character} relating to invariance under space-time displacements and
uniform motion. The best known {\it local} invariance operations   (i)
relate different coordinate systems to one other by such {\it geometric}
deformations  that leave the (local) metric invariant and (ii) appear as
{\it local} transformations that link different gauge theory
representations to one another.  It was recently realized that probing
{\it non-local} (topological) structures uncovers new invariance
principles with physical (and experimental) consequences. Profound {\it
particle-wave} type dualities link the seemingly different local and
non-local structures.  

Understanding the thermodynamic phases of matter via symmetry principles
enables characterization by universal behaviors as in Landau's theory of
phase transitions \cite{Landau}.  In this theory, the order parameter(s)
of the system relate to probing  its {\it local} structure. A
non-vanishing  order parameter, encoding the breaking of a symmetry,
defines the  {\it ordered} state while the restoration of that symmetry
signals the transition to a {\it disordered} state. A new paradigm,
Topological Quantum Order (TQO), extends the Landau symmetry-breaking
framework \cite{wenbook}.  In essence, this new order is associated with
robustness against {\it local} perturbations, and hence cannot be
described in principle by {\it local} order parameters. Thus, the
underlying order remains {\it hidden} to ordinary {\it local} probes. 
Indeed, this new order exhibits {\it non-local} correlations that
potentially lead to  novel physical consequences. Interest in TQO is
further enhanced by the prospect of using it to engineer fault tolerant
hardware for quantum computation \cite{kitaev}. A main objective of the
present manuscript is to introduce a framework based on symmetry to
study this new kind of order of matter.  Key questions concern the
physical organizing principles underlying such an order, how TQO does
manifest, and how to mathematically characterize it.  All of the
symmetry based  results that we will outline here apply also to systems
displaying  {\em emergent} symmetries \cite{GJW}. The latter are not
actual invariances of the system but {\it emerge} as exact symmetries
at low energies.  

Several inter-related concepts are associated with TQO such as
symmetry, degeneracy, fractionalization of quantum numbers, and maximal
string/brane correlations ({\it non-local} order).  It is important to
determine what is needed for a system to display this order. While
canonical examples such as the Fractional Quantum  Hall liquids exist,
there is no unambiguous definition of TQO. We start by defining this
order following Ref. \cite{kitaev} and show relations between these
{\it different} concepts, establishing  the equivalence between some
and more lax relations amongst others. Most importantly, we suggest a
symmetry principle for TQO. In particular, we 
\newline
(i) prove that systems with $d$-dimensional (with $d=1,2$) 
Gauge-Like Symmetries exhibit TQO; 
\newline  
(ii) analyze the resulting conservation laws and emerging topological
terms;
\newline
(iii) affirm that the energy spectrum on its own is insufficient for
establishing the existence  of this order (the {\it devil} is in the
state itself);  
\newline
(iv) suggest that while, fractionalization, string/brane-type
correlators, and entanglement entropy are important concepts they are
not always sufficient conditions for TQO; 
\newline
(v) sketch a general algorithm for constructing string/brane
correlators.

Our goal is to provide a unifying framework which will identify  new
physical models with TQO. A detailed derivation of the reported results  
appears in Ref. \cite{Longversion}.


We focus on quantum lattice systems (and their continuum extension) 
having  $N_s= L^{D}$ sites, with $L$ the number of sites along each
space direction, and $D$ the  dimensionality of the lattice $\Lambda$.
Associated with each lattice site (or mode, or bond, or plaquette,
etc.) ${\bf i} \in \Z^{N_s}$ there is a Hilbert space ${\cal H}_{\bf
i}$ of finite dimension ${\cal D}$. The full Hilbert space is the
tensor product of the local state spaces, ${\cal H} = \bigotimes_{\bf
i} {\cal H}_{\bf i}$, in the case of distinguishable subsystems, or a
proper subspace in the case of indistinguishable ones. Statements about
local order, TQO, fractionalization, entanglement, etc., are {\it
relative} to the particular decomposition used to describe the physical
system. The physically motivated natural {\it local language}
\cite{GJW} is often utilized. The definition of TQO requires a set of
$N$ orthonormal ground states $\{|g_{\alpha} \rangle\}_{\alpha
=1,\cdots, N}$ with a gap to excited states. These states should be
topologically distinct with no bounded operator $V$ with compact
support connecting any pair of such states nor, equivalently, can such
quasi-local operators be used to distinguish between different ground
states. Specifically, TQO exists if and only if for any quasi-local
operator $V$,
\begin{eqnarray}
\langle g_{\alpha} | V | g_{\beta} \rangle = v \ \delta_{\alpha \beta} +
c ,
\label{def.}
\end{eqnarray}
where $v$ is a constant and $c$ is a correction that it is either zero
or vanishes in  the thermodynamic limit. That is, the ground states 
locally look identical but not globally.  We  will also examine a finite
temperature ($T >0$) extension for the diagonal elements of Eq.
(\ref{def.}),
\begin{eqnarray}
\langle V \rangle_{\alpha}  \equiv \tr(\rho_{\alpha} V)= v + c
~~~~(\mbox{independent of~} \alpha)
\label{vt} ,
\end{eqnarray} 
with $\rho_{\alpha} = \exp[-H_{\alpha}/(k_{B} T)]$  a density matrix
corresponding to the Hamiltonian $H$ when augmented by an infinitesimal
operator $\alpha$. At $T=0$, a particular choice of such operators can
be constructed to  favor the state  $|g_{\alpha} \rangle$.  We consider
the most general case of arbitrary  infinitesimal operators. Such
infinitesimal operators may be non-local, e.g.,  may couple to the toric
cycle operators \cite{Longversion} in Kitaev's toric code model
\cite{kitaev}.  A system displays finite-$T$ TQO if it satisfies both
Eqs. (\ref{def.}), and  (\ref{vt}). Those equations describe
mathematically the fact that physical states cannot be distinguished by
local measurements. The only way to extract non-trivial information is
to perform a non-local measurement. If the expectation value of any
quasi-local operator $V$ is independent of  topological constraints that
favor the state $|g_\alpha \rangle$, or a collection of such states in a
given topological sector, then the system exhibits TQO:  the information
encoded in the states is unaccessible (and protected) from any
quasi-local perturbation. Such a topological sector may be defined by
Hopf invariants, total domain wall parity, etc.

Note that the conditions (\ref{def.}) and (\ref{vt}) enable TQO only if
the ground state sector contains, at least, two orthogonal states. 
According to these conditions, systems with non-degenerate ground states
(such as Integer Quantum Hall liquids) do not support TQO. [In fact,
such non-degenerate systems exhibit no spontaneous symmetry breaking and
generally also do not support any Landau orders.]  As we showed, {\em a
necessary (but not sufficient) criterion for robust topological quantum
memories} is directly related to the conditions of Eqs. (\ref{def.},
\ref{vt}) for TQO \cite{usnew}. That this criterion is necessary but
insufficient is exemplified by Kitaev's toric code model \cite{kitaev}
which as  we discuss here and in far more detail elsewhere
\cite{usnew}   exhibits a finite (system size independent)
autocorrelation time.  This criterion for robust quantum memories
applies to systems with both degenerate and non-degenerate ground states
(see Appendix \ref{app1}). It shows that non-degenerate systems are
trivial from another viewpoint as well.  Quantum computing applications
of TQO rely on the non-commutativity of different braiding operations.
The presence of non-commuting symmetries ensures  a {\em topological
degeneracy} in the ground state sector.  

\begin{figure}
\centering
\includegraphics[scale=0.7]{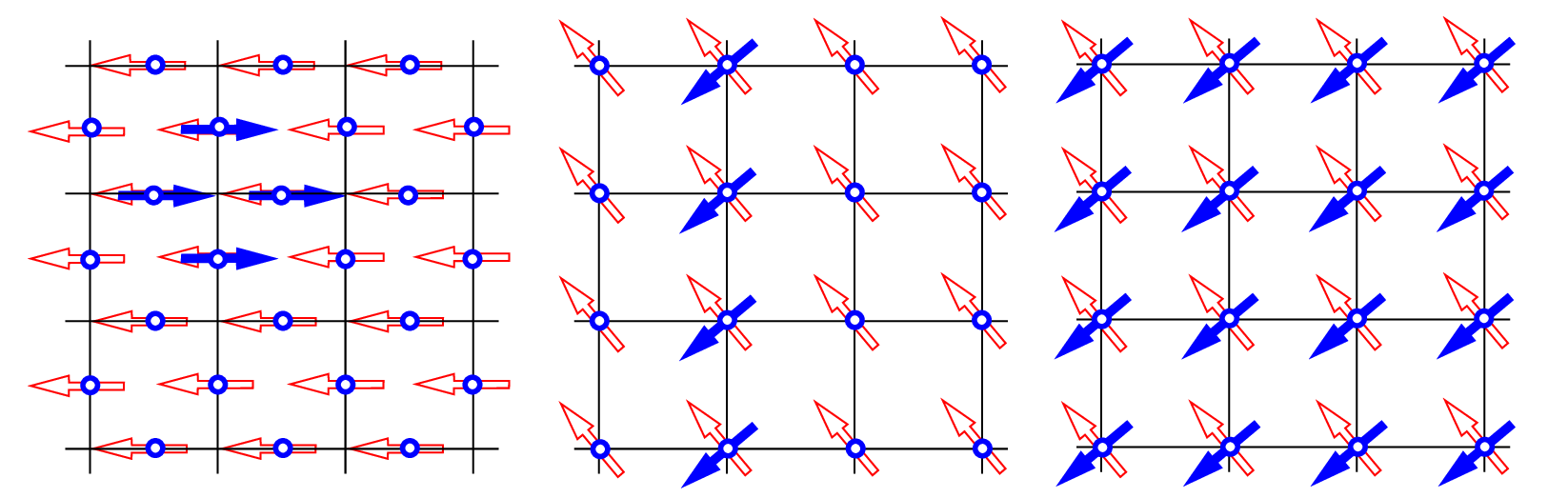}
\caption{Schematics of the interactions and symmetries involved in 
three $D=2$ examples.  See (a-c) of text. The left panel represents (a)
an Ising gauge theory with local ($d=0$) symmetries.  The middle panel
represents (b) an orbital compass model with $d=1$ symmetries; here the
symmetry operations span lines.  The right panel depicts (c) an XY model
with $d=2$ symmetries; the symmetry here  spans the entire $D=2$
dimensional plane.}
\label{examples}
\end{figure}

To put these definitions in perspective, consider the typical example
of local order: the spin $s=1/2$  Heisenberg ferromagnet. The
ferromagnet has a continuum of ground states $|g_{\hat{n}} \rangle$
with a magnetization in the direction $\hat{n}$. In the thermodynamic
limit, these states become orthogonal to one another 
\begin{eqnarray}
\langle g_{\hat{n}} | g_{\hat{\ell}} \rangle = \Big( \frac{1}{2}
(1+ \hat{n} \cdot \hat{\ell}) \Big)^{N_{s}/2} \stackrel{
N_{s} \to \infty}{\longrightarrow}  0.
\end{eqnarray}
However, in this case, e.g. the local spin operator $V=S_{\bi}^{z}$
may distinguish between different ground states: 
\begin{eqnarray}
\langle g_{\hat{n}}| V | g_{\hat{n}}\rangle = \frac{\hbar}{2}(\hat{n} \cdot
\hat{e}_{z})\neq \langle g_{\hat{\ell}}| V | g_{\hat{\ell}}\rangle ,
\end{eqnarray}
with $\hat{e}_{z}$ a unit vector along the $z$ axis. Thus, Landau type
systems with local order parameters do not even satisfy the $T=0$
condition of Eq. (\ref{def.}). Similarly, in three spatial  dimensions,
at finite temperatures, 
\begin{eqnarray}
\langle  V  \rangle_{\hat{n}} \neq  \langle V \rangle_{\hat{\ell}} .
\label{vab}
\end{eqnarray}
A $d$-dimensional Gauge-Like Symmetry of a theory is a group of
symmetry transformations ${\cal{G}}_d$ such that the minimal  set of
fields $\phi_{\bf i}$ changed by the group operations is located on a 
$d$-dimensional subset  ${\cal C} \subset \Lambda$ of the complete
$D$-dimensional lattice ($d \le D$). These transformations can be 
expressed as \cite{BN}:
${U}_{{\bl\bk}} = \prod_{{\bf i} \in {\cal C}_{\bl}} {\bf g}_{{\bf
i}{\bk}}$,
where ${\cal C}_{\bl}$ denotes the (external) spatial subregion $\bl$,
$\Lambda= \bigcup_{\bl} {\cal C}_{\bl}$, and $\bk$ labels the
(internal) group generators.  [The extension of this definition to the
continuum is straightforward.] Gauge (local) symmetries correspond to
$d=0$, while in global symmetries the region influenced by the symmetry
operation is the full $d=D$-dimensional volume of the system. These
symmetries may be Abelian or non-Abelian.  An illustration is provided
in Fig.  \ref{examples} which describes three different $D=2$ systems
with symmetries of dimensions $d=0,1,2$:

(a) The Ising gauge theory whose Hamiltonian is $H = - K \sum_{p}
\sigma^{z}_{ij}  \sigma^{z}_{jk} \sigma^{z}_{kl} \sigma^{z}_{li}$. The
sum is over all elementary squares ({\it plaquettes}) $p$,  and on each
bond $(ij)$ we have an $s=1/2$ {\it spin} $\hbar \vec{\sigma}_{ij}/2$
($\vec{\sigma}_{ij}$ represents the triad of Pauli matrices). In this
system, we have the local gauge symmetries $G_{i} = \prod_{s}
\sigma^{x}_{is}$ with the sites $\{s\}$ being the nearest-neighbors of
$i$.  For any lattice site, $[G_{i}, H]=0$. $\{G_{i}\}$ involve a
finite number of local operators and thus constitute $d=0$ symmetries.

\begin{figure}
\centering
\includegraphics[width=0.975\columnwidth]{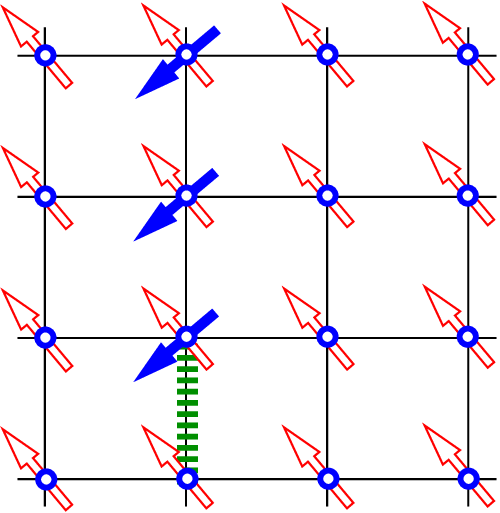}
\caption{(Color online.) The physical engine behind our theorem are {\underline{{\it topological 
defects}}}. For instance, introducing a (one-dimensional) soliton into a
general (anisotropic) orbital compass model state such as depicted in
panel (b) of Fig.~\ref{examples} leads to only a finite energy cost.
This   penalty is depicted here by a single energetic bond (dashed
line). The energy-entropy balance associated with such $d=1$ Ising type
domain walls is the same as that in a $D=1$ Ising system. At finite
temperature, entropic contributions  overwhelm energy penalties and
no  local order is possible. Order is manifest in non-local quantities
associated with topological defects. Similar results occur in other
systems with low-dimensional Gauge-Like Symmetries.}
\label{soliton}
\end{figure}

(b) The anisotropic $s=1/2$ orbital compass model \cite{ocm} 
\begin{eqnarray} H = -
\sum_{i} [J_{x}\sigma^{x}_{i} \sigma^{x}_{i+ \hat{e}_{x}} + J_{z}
\sigma^{z}_{i} \sigma^{z}_{i+\hat{e}_{z}}],
\label{ocmeq}
\end{eqnarray} 
emulating the direction dependent interactions of electronic orbitals. 
In this model, exchange interactions involving  the $x$ component of the
spin occur only along the spatial $x$ direction of the lattice. Similar
spatial direction dependent spin exchange interactions appear for the
$z$ components of the spin. Apart from a global reflection which only
appears for the isotropic  point $J_{x} = J_{z}$ (and which may be
broken at low $T$),  the  orbital compass model has the following
symmetries $O^{\alpha} =  \prod_{j \in C_{\alpha}} i\sigma_{j}^{\alpha}$
for $\alpha = x,z$.  $C_{\alpha}$ denotes any line orthogonal to the
$\hat{e}_{\alpha}$ axis. On a torus, these operators are defined along
toric cycles. As $O^{\alpha}$ involves ${\cal{O}}(L^{1})$ sites, they
constitute $d=1$ symmetries. As we will elaborate on, these $d=1$ 
symmetries cannot be broken at finite $T$. [See also
Fig.~\ref{soliton}.] It is worth emphasizing that global symmetries such
as {\it time reversal} \cite{Longversion} may be seen as composites of the
minimal (and more fundamental) $d=1$  symmetries of $O^{\alpha}$  
\cite{timereversal}.

(c) The $s=1/2$ XY ferromagnet $H = - J \sum_{\langle ij \rangle} 
[\sigma^{x}_{i} \sigma^{x}_{j} + \sigma^{z}_{i} \sigma^{z}_{j}]$ has
continuous $d=D=2$ symmetry operators: $U(\theta)=  \prod_{j} \exp[
- (i/2) \theta  \sigma^{y}_{j}]$. There are ${\cal{O}}(L^2)$ local
operators in the product which defines the $d=2$ operator $U(\theta)$. 

\onecolumngrid
\twocolumngrid

The group symmetry operators and topological defects can be written in
terms of unitary operators. In the continumm limit, the group elements
of $d=1$ Gauge-Like Symmetries are path-ordered (${\cal P}$) products
\begin{equation}
U = {\cal P} e^{ i \oint_C \vec{A}  \cdot \vec{dr}} ,
\label{u1}
\end{equation} 
where $C$ is a closed path in configuration space and $\vec{A}$ is the
corresponding connection. On the lattice this expression is replaced
by  an equivalent discrete sum, and  $C$ is a closed path on the
lattice. $U$ is an Aharonov-Bohm phase \cite{AB} or a Wilson loop
\cite{kogut}. For instance, for the orbital compass model [(b) above], 
the $d=1$ group elements can be written as
\begin{equation}
U=O^x= \prod_{j \in C_x} i\sigma^{x}_{j}  =e^{i \frac{\pi}{2}
\sum_{j \in C_x} \sigma^{x}_{j}}.
\label{u2}
\end{equation}
A defect creating operator in such a system [see e.g. the defect in
Fig. \ref{soliton}] is
\begin{eqnarray}
D_{+}=  {\cal P} e^{ i \int_{C_{+}} \vec{A}  \cdot \vec{dr}},
\end{eqnarray}
or an exponentiated discretized sum.  Here, $C_{+}$ is an open contour.
For instance, the defect depicted in Fig.~\ref{soliton} is generated by
\begin{eqnarray} 
D_{+} = e^{i \frac{\pi}{2} \sum_{j \in C_{+}} \sigma^{x}_{j}} .
\end{eqnarray}
We can interpret $D_{+}$ as the creation of a defect-antidefect pair
and  the displacement of each member of the pair to the opposite
endpoints of $C_{+}$.  The $d=1$ Gauge-Like Symmetry operators linking
different ground states ($U$ of Eqs. (\ref{u1}, \ref{u2})) correspond
to the displacement  of a defect-antidefect pair along a toric cycle.
Formally, this is similar to the quasiparticle-quasihole pair creation
operator linking different ground states  in the Fractional Quantum
Hall Effect \cite{frank}.  Topological concepts such as {\it monodromy}
and {\it homotopy} can be realized by $d=1$ symmetry operators.  More
complicated topological properties are associated with $d=2$
symmetries. 

What are the physical consequences for a system with a symmetry group
${\cal{G}}_d$? Most importantly, these symmetries lead to an effective {\em
dimensional reduction} in several characteristics. 

The great insight behind the Landau theory of phase transitions lies in
the intimate relation between the  spontaneous breaking of a symmetry in
the Hamiltonian and the  appearance of a new ordered phase of matter
characterized by an order parameter. Can we spontaneously break 
$d$-dimensional Gauge-Like Symmetries? An important key for addressing
this question is given by an inequality. This inequality states that the
absolute values of quantities not invariant under ${\cal{G}}_d$ are
bounded from above by the expectation values that they attain in a
$d$-dimensional Hamiltonian $\bar{H}$  (or corresponding action
$\bar{S}$) which is globally invariant under ${\cal{G}}_d$ and preserves
the range of the interactions of the original system \cite{BN}. [See the
Appendix of Ref. \cite{Longversion} for a review of this inequality.] As
the expectation values of local observables vanish in low-$d$ systems,
this bound strictly forbids spontaneous symmetry breaking of 
non-${\cal{G}}_d$ invariant local quantities in systems with
interactions of finite range and strength whenever $d=0$ (Elitzur's
theorem \cite{Elitzur}), $d=1$ for both discrete [see Fig. 
\ref{soliton}] and continuous ${\cal{G}}_d$, and (as a consequence of
the Mermin-Wagner-Coleman theorem \cite{Mermin}) whenever $d=2$ for
continuous symmetries \cite{BN}. Discrete $d=2$ symmetries may be broken
(e.g. the finite-$T$ transition of the $D=2$ Ising model, and the $d=2$
Ising Gauge-Like Symmetries of $D=3$ orbital compass systems).  In the
presence of a finite gap in a system with continuous $d < 2$ symmetries,
spontaneous symmetry breaking  is forbidden even at $T=0$ \cite{BN}. The
absence of spontaneous symmetry breaking of $d$-dimensional Gauge-Like
Symmetries is due to the presence of low-($d$) dimensional topological
defects such as: domain walls/solitons in systems with  $d=1$ discrete
symmetries, vortices in systems with $d=2$ $U(1)$ symmetries, hedgehogs
for $d=2$ $SU(2)$ symmetries. Transitions and crossovers can only be
discerned by quasi-local symmetry invariant quantities (e.g.,
Wilson-like loops) or, by probing global topological  properties (e.g.
percolation in  lattice gauge theories \cite{percolation}). Extending
the bound of \cite{BN} to $T=0$, we  now find that if $T=0$ spontaneous
symmetry breaking is precluded in systems with $d$-dimensional
Gauge-Like Symmetries,  then spontaneous symmetry breaking of 
quantities not invariant under exact or $T=0$ {\em emergent}
$d$-dimensional Gauge-Like Symmetries cannot occur as well.  Exact
symmetries refer to $[U,H]=0$;  in {\em emergent} symmetries \cite{GJW}
unitary operators $ U \in {\cal{G}}_{\sf emergent}$ are not bona fide
symmetries ($[U,H] \neq 0$) yet  become exact at low energies: when
applied to any ground state, the resultant state must also  reside in
the ground state manifold,  $U| g_{\alpha} \rangle = \sum_{\beta}
u_{\alpha  \beta} | g_{\beta} \rangle$. 
In gapped systems, $T=0$ spontaneous symmetry breaking of $d < 2$
continuous symmetries is prohibited. In particular, it can be seen
\cite{Longversion} that even {\em emergent} symmetries that appear only
within the ground state sector in systems with a spectral gap preclude
the  appearance of spontaneous symmetry breaking. The proof of this
statement relies on (i) extensions of the Mermin-Wagner-Coleman theorem
to gapped systems with continuous symmetries at $T=0$ \cite{assa} in
dimensions smaller than two, and on  (ii) an application of general
bounds, derived in \cite{BN}, to  these problems  \cite{Longversion}. 

Symmetries generally imply the existence  of conservation laws and
topological charges, with associated (local) continuity equations when
the group of symmetries is continuous. Moreover, if the continuous
symmetry forms a  gauge group, then an additional local Gauss' law is
satisfied by the conserved currents. We find that systems with
$d$-dimensional Gauge-Like Symmetries  lead to conservation laws within
$d$-dimensional regions. To illustrate this, consider the rotational
non-invariant Euclidean Lagrangian density of a complex field
$\vec{\phi}(\vec{x})=(\phi_1(\vec{x}),
\phi_2(\vec{x}),\phi_3(\vec{x}))$ ($\vec{x}=(x_1,x_2,x_3)$):
${\cal{L}} = \frac{1}{2} \sum_{\mu} |\partial_{\mu} \phi_{\mu}|^{2} 
+ \frac{1}{2} \sum_{\mu} |\partial_{\tau} \phi_{\mu}|^{2} +W(\phi_\mu) $,
with $W(\phi_\mu)= u(\sum_{\mu} |\phi_{\mu}|^{2} )^{2} - \frac{1}{2}
\sum_{\mu} m^{2}(|\phi_{\mu}|^{2})$ and $\mu,\nu=1,2,3$ are the spatial
directions and $\tau$ is the imaginary time. ${\cal L}$  displays the
continuous $d=1$ symmetries $\phi_{\mu} \to e^{i
\psi_{\mu}(\{x_{\nu}\}_{\nu\neq\mu})}  \phi_{\mu}$.  The conserved $d=1$
Noether currents are given by  $j_{\mu \nu}= i [\phi_{\mu}^{*}
\partial_{\nu}  \phi_{\mu} - (\partial_{\nu}  \phi_{\mu}^{*})
\phi_{\mu}]$, which satisfy $d=1$ conservation laws $[\partial_{\nu}
j_{\mu \nu} + \partial_{\tau} j_{\mu \tau}] =0$ (with no summation over
repeated indices implicit). Such $d$-dimensional Gauge-Like Symmetries 
may lead to a conservation law for each line associated with a fixed
value of all coordinates $x_{\nu \neq \mu}$ relating to the topological
charge  $Q_{\mu}(\{x_{\nu \neq \mu}\}) = \int dx_{\mu} ~j_{\mu
\tau}(\vec{x})$ (no summation over the repeated index $\mu$).

A similar consequence of $d$-dimensional Gauge-Like Symmetries   is
that topological terms appearing in $d+1$-dimensional theories also
appear in higher $D+1$-dimensional systems ($D>d$). These topological
terms appear in actions $\bar{S}$ [see Appendix or Ref. \cite{BN}]
which may be used to bound (both from above and below) expectation
values of quantities that are not invariant under the  $d$-dimensional
Gauge-Like Symmetries.  For instance, in the isotropic $D=2$ (or 2+1)
dimensional  general spin $t_{2g}$ Kugel-Khomskii model \cite{KK,
Longversion},  of exchange constant $J>0$, the corresponding continuum
Euclidean action used to bound the $d=1$ symmetry non-invariant
quantities is of the 1+1 form
$\bar{S} = \frac{1}{2g} \int dx d \tau  \Big[ \frac{1}{v_{s}}
(\partial_{\tau} \hat{m})^{2} - v_{s} (\partial_{x} \hat{m})^{2} \Big] 
+ i \theta {\cal{Q}}  + S_{{\sf tr}},~~  {\cal{Q}} = \frac{1}{4 \pi}
\int dx d\tau \Big[ \hat{m} \cdot (\partial_{\tau} \hat{m} \times
\partial_{x} \hat{m}) \Big]$.
Here, as in the non-linear-$\sigma$ model of a spin-$s$ chain, $\hat{m}$
a normalized slowly varying  staggered field, $g=2/s$, the spin wave
velocity  $v_{s} = 2 Js$, $\theta = 2 \pi s$, and $S_{{\sf tr}}$ a {\it
transverse-field} action term which does  not act on the spin degrees of
freedom along a given chain.  ${\cal{Q}}$ is the Pontryagin index 
corresponding to the mapping between the two-dimensional space-time
$(x,\tau)$ plane and the two-sphere on which $\hat{m}$ resides. This 1+1
dimensional topological term appears  in the 2+1 dimensional
Kugel-Khomskii system even for arbitrary large positive coupling $J$.
This, in turn, places bounds on the spin correlations enabling us to
predict, for instance, that in $D=2$ integer-spin $t_{2g}$
Kugel-Khomskii systems, a finite correlation length exists. 

Is there a relation between $d$-dimensional Gauge-Like Symmetries and
the existence of fractional quantum numbers? Symmetries do not
necessarily mandate the existence of degeneracies at all energies.
Degeneracies, however, imply the existence of symmetries which effect
general unitary transformations within the degenerate manifold and act
as the identity operator outside it.  We note that the most general
symmetry of any Hamiltonian is given by a direct product of the form
$\bigotimes_{l} SU(N_{l})$ with $N_{l}$ being the degeneracy of the
$l$-th eigenvalue of the Hamiltonian. Any projection of a
symmetry operator onto the ground state sector must constitute an
element of $SU(N_{g})$ with $N_{g}$ the number of ground states. Any
emergent symmetry must also act as an $SU(N_{g})$ operator. In this way,
these unitary transformations may act as exact Gauge-Like Symmetries.
The existence of such unitary symmetry operators (generally a subset of
$SU(N)$) allows for fractional charge (such as the ``triality'' for the
$SU(3)$ group of quantum chromodynamics or ``$N$-ality" of the $SU(N)$
symmetry group). This then suggests that degeneracy allows for a
fractionalization defined by the center of the symmetry group (which may
be $SU(N)$ or any of its subgroups/quotient groups).  The ($m$-)rized 
Peierls chains constitute a typical example of a system with universal
($m-$independent) symmetry operators \cite{Longversion}, where
fractional charge quantized in units of  $e^{*} = e/m$ with  $e$ the
electronic charge is known to occur  \cite{fraction}.  The bounds above
can be generalized to apply to these symmetries. Different Peierls chain
ground states break  discrete $d=1$ symmetries (violating Eq.
(\ref{def.}) in this system with  fractionalization), meaning that
fractionalization may occur in systems  with no TQO. The fermion number
$N_{f}$  in the Peierls chain and related Dirac theories  is an integral
over spectral functions \cite{NS};  the fractional portion of $N_{f}$
stems from  soliton contributions invariant under local background
deformations.


When the bound of \cite{BN} is next applied to correlators and spectral
functions, it implies the absence of quasi-particle  excitations in
many instances \cite{NBF}.
Here we elaborate on this: The bound of \cite{BN} mandates that the
absolute values of non-symmetry invariant  correlators $|G| \equiv
|\sum_{\Omega_{{\bf j}}} a_{\Omega_{{\bf j}}} \langle \prod_{{\bf i}
\in \Omega_{{\bf j}}}  \phi_{{\bf i}} \rangle|$ with $\Omega_{{\bf j}}
\subset {\cal C}_{{\bf j}}$, and  $\{a_{\Omega_{{\bf j}}}\}$ 
$c$-numbers, are bounded from above (and from below for $G\geq0$ (e.g.
that corresponding to $\langle |\phi({\bf k}, \omega)|^{2} \rangle$))
by absolute values of the  same correlators $|G|$ in a  $d$-dimensional
system defined by ${\cal C}_\bj$. In particular, when ${\bf k}$ lies in
a lower dimensional region where the $d$-dimensional Gauge-Like
Symmetry is present (e.g. for ${\bf{k}}$ parallel to the
${\hat{e}}_{x}$  or ${\hat{e}}_{z}$ axis of the orbital compass model of
Eq. (\ref{ocmeq})), the coefficients $\{a_{\Omega_{{\bf j}}}\}$ can be
chosen to give the Fourier transformed  pair-correlation functions.
This provides bounds on viable quasi-particle weights and establishes
the absence of quasi-particle excitations in many cases. In
high-dimensional interacting systems, retarded correlators $G$
generally exhibit a resonant (quasi-particle) contribution. 
$G= G_{{\sf res}}({\bf k}, \omega)  + G_{{\sf non-res}}({\bf k},
\omega)$ with $G_{{\sf res}}({\bf k}, \omega)  = \frac{Z_{{\bf
k}}}{\omega - \epsilon_{{\bf k}} + i 0^{+}}$. In formal terms, in
high-dimensional systems, the interactions amongst otherwise bare free
particles simply ``renormalize" those free particles. The interactions
replaced free particles by new quasi-particles which are for all
purposes just bare  particles with new renormalized weights and
effective parameters. This quasi-particle behavior is captured by the
poles of the retarded correlator. This is not so in low-dimensional
systems \cite{vns}. In low dimensions, particle-particle interactions
can lead to a dramatic change of the system. The system no longer has
quasi-particle type behavior: the quasi-particle weight $Z_{{\bf k}}
\to 0$ and the poles of $G$ are often replaced by weaker branch cut
behavior. Fractionalization is associated with the  disappearance of
quasi-particles with sharp dispersion relations. If the momentum ${\bf
k}$ lies in a lower $d$-dimensional region ${\cal{C}}_{\bf j}$  and if
no quasi-particle resonant terms appear in the corresponding
lower-dimensional spectral functions in the presence of non-symmetry
breaking fields then the upper bound \cite{BN} on the correlator $|G|$ 
(and on related quasi-particle weights given by $Z_{{\bf k}} =
\lim_{\omega \to \epsilon_{{\bf k}}} (\omega - \epsilon_{{\bf k}})
G({\bf k}, \omega)$) of symmetry non-invariant quantities mandates the
absence of normal quasi-particles. ``Non-symmetry breaking fields"
refer to fields outside the region where the $d$-dimensional Gauge-Like
symmetry is present - i.e. the fields $\{\phi_{\bf i} \not \in
{\cal{C}}\}$. Using the bound of \cite{BN} 
for $Z_{{\bf k}} = \lim_{\omega \to \epsilon_{{\bf k}}}
(\omega - \epsilon_{{\bf k}}) G({\bf k}, \omega)$ we see that if
fractionalization occurs in the lower-dimensional system ($Z_{\bf k}
=0$)  then its higher-dimensional realization follows. These external
fields do not (by the very presence of the $d$-dimensional Gauge-Like
Symmetry: $[H,U]=0$)  break the symmetry. Fractionalized excitations
propagate  in $d_{s} = D-d$ dimensional regions \cite{NBF}. Once again,
it is important to stress that as bounds concerning correlators in
low-dimensional gapped systems may be extended to zero temperature, the
bound of \cite{BN} implies the absence of quasi-particle excitations
also in cases in which continuous $d <2$ Gauge-like symmetries only
emerge within the low energy sector.

Our central claim is that in all known systems with TQO,
$d$-dimensional Gauge-Like Symmetries  are present. Old examples
include: Fractional Quantum Hall systems, $\Z_{2}$ lattice gauge
theories, the toric code model \cite{kitaev}, and others.  In all cases
of TQO, we may express known {\it topological} symmetry operators as
general low-dimensional $d \le 2$  Gauge-Like Symmetries  (e.g. in the
toric code model, there are $d=1$ symmetry operators  spanning toric
cycles). These symmetries allow for freely-propagating decoupled
$d$-dimensional topological defects  (or instantons in $(d+1)$
dimensions of Euclidean space-time) which eradicate local order. These 
defects enforce TQO. We now state a central result: 

{\bf{Theorem.}} {\em Consider a system with interactions of finite
range and strength that satisfies Eq. (\ref{def.}). If all ground
states may be linked by  discrete $d \le 1$ or by  continuous $d \le
2$-dimensional Gauge-Like Symmetries  $U \in {\cal{G}_{\it d}}$, then 
the system displays finite-$T$ {\rm TQO}. } 

\proof  
Let us start by decomposing $V = V_{0} + V_{\perp}$. Here, $V_{0}$ is 
the part transforming as a singlet under ${\cal G}_d$ i.e.,
$[U,V_{0}]=0$.  To prove the finite-temperature relation of  Eq.
(\ref{vt}),   we write the expectation values over a complete  set of
orthonormal energy states $\{ |a \rangle \}$,
\begin{eqnarray}
\!\!\!\!\!\!\!\! \langle V_{0}  \rangle_{\alpha}  &=& \lim_{h \to
0^{+}}  \frac{\sum_{a} \langle a| V_{0}| a \rangle  e^{-(E_{a} +
\phi^{a}_{\alpha}
h)/k_B T}}{\sum_{a} e^{-(E_{a} + \phi^{a}_{\alpha}
h)/k_B T}} \nonumber \\
&=& \lim_{h \to 0^{+}} \frac{ \sum_{a} \langle a| U^{\dagger} V_{0} U | a
\rangle  e^{-(E_{a} + \phi^{a}_{U  \beta}
h)/k_B T}}{\sum_{a} e^{-(E_{a}  + \phi^{a}_{U \beta}
h)/k_B T}} \nonumber \\
&=& \lim_{h \to 0^{+}}  \frac{\sum_{b} \langle b| V_{0}| b \rangle
e^{-(E_{b} +  \phi^{b}_{\beta}
h)/k_B T}} {\sum_{b}
e^{-(E_{b} + \phi^{b}_{\beta}
h)/k_B T}} = \langle V_{0}
\rangle_{\beta}.
\label{long}
\end{eqnarray}

We invoked $U|a \rangle \equiv |b \rangle,$ and $E_{a} = E_{b}$
following from $[U,H]=0$. The term $\phi^{a}_{\alpha}$ is the
expectation value of the infinitesimal operator $\alpha$ in the  state
$| a \rangle$. That is,  $\langle a |\alpha |a  \rangle  \equiv h
\phi_{\alpha}^{a}$ with $h = \|\alpha\|$ the operator norm of $\alpha$.
In the above derivation,  $\phi^{a}_{\alpha} = \phi^{Ua}_{U \alpha}  =
\phi^{b}_{\beta}$ with $\beta \equiv U^{\dagger} \alpha U$. As
$V_{\perp}$ is not invariant  under ${\cal{G}_{\it d}}$ ($[U,V_{\perp}] 
\neq 0$) the theorem of \cite{BN} implies that $\langle V_{\perp}
\rangle_\alpha =0$,   i.e., spontaneous symmetry breaking is precluded
in systems with low-dimensional Gauge-Like Symmetries. [See
Fig.~\ref{soliton}.] Equation (\ref{long}) is valid whenever $[U,V_{0}]
= 0$ for {\it any} symmetry $U$. However,  $[U,V_{\perp}] \neq 0$
implies $\langle V_{\perp}  \rangle_\alpha =0$ only if $U$ is a
low-dimensional Gauge Like Symmetry. In systems in which not all ground
state pairs can be linked by the use of low-dimensional Gauge-Like
Symmetries  $U \in {\cal{G}}_{d}$ ($U| g_{\alpha} \rangle = | g_{\beta}
\rangle$), finite-temperature  spontaneous symmetry breaking may occur.
We conclude this proof with two remarks  \cite{Longversion}: (i) $T=0$
TQO holds whenever all ground states may be linked by (exact or
emergent) continuous  $d < 2$ Gauge-Like Symmetries in gapped systems.
(ii) In many systems (whether gapped or gapless),  $T=0$ TQO states may
be constructed by  employing Wigner-Eckart type selection rules for  $d
\ge 1$ Gauge-Like Symmetries. It is important to emphasize that the
$T=0$ conditions of TQO [Eq.~(\ref{def.})] can be violated for some
ground states of  theories with local ($d=0$) symmetries. In such
theories,  ground states can be linked by $d=0$ operators; choosing $V$
to be such a quasi-local ($d=0$) operator the TQO requirement of
Eq.~(\ref{def.}) is violated.

What physical quantity characterizes TQO? We will see that neither the
spectrum of $H$, nor the entanglement entropy \cite{entropy}, nor
string/brane correlations are sufficient criteria. That the spectrum, on
its own, is insufficient is established by  counter-examples.  For
example, Kitaev's  \cite{kitaev} and Wen's models \cite{wen_plaq} are
equivalent and have a spectrum  identical to that of two uncoupled Ising
spin chains with nearest-neighbor interactions  (which we have shown by
applying a duality mapping \cite{Longversion}).  While  Kitaev's  and
Wen's models display TQO, the Ising chains do not. By duality,   the
exchange statistics of defects  (the so-called magnetic and electric
charges) of Kitaev's model that reflect the change incurred while moving
one  excitation completely around another (i.e., {\it braiding}), can be
similarly cast as operator relations that may be represented on two
Ising chains. The mapping of the braiding operations is, in general,
path dependent.  The electric and magnetic particles of Kitaev's model
map onto domain walls in the two respective Ising chains. The operator
products corresponding to full braiding operators of defects in Kitaev's
model can be cast as string products of the spins on the Ising chains
when domain walls are present.  Our duality mappings to Ising chains
demonstrate that despite  the spectral gap in these systems, the toric
operator expectation values may vanish once thermal fluctuations are
present (a concept we coined {\it thermal fragility} \cite{Longversion,
usnew}).  The Ising chains have a zero temperature transition. In
Kitaev's toric code model \cite{kitaev}, an explicit calculation of the
temporal autocorrelation function amongst the $d=1$ {\it topological
invariants} using these mappings shows that the autocorrelation time is
finite at any positive temperature \cite{usnew}. The vanishing
expectation values of the toric code operator reflect a finite
autocorrelation time.

The autocorrelation time $\tau$ of the toric code operators
(e.g., $\langle X_{i}(0) X_{i}(t) \rangle$) at an inverse
temperature $\beta$ is 
is identical to that in Ising chains. For Glauber type
dynamics, we thus find that the autocorrelation time of the toric code
operators \cite{usnew},
\begin{eqnarray}
\tau = \frac{\mbox{const}}{1-\tanh 2 \beta}.
\label{glauberKitaev}
\end{eqnarray} 
For periodic boundary conditions and in the asymptotic low-temperature
limit, this autocorrelation time is related to the spectral gap $\Delta$
between the ground state energy and the energy of the next excited state
by $\tau \to_{\beta \to \infty} \exp(\beta \Delta)$.  Thus, although
finite, the autocorrelation time can  be made very large if temperatures
far below the  gap may be achieved.  In particular, we discussed the
detailed physical meaning of this result in \cite{usnew}.  It
illustrates that realizing finite-$T$ fault tolerant quantum memories is
a subtle problem. Finite-$T$ TQO  (Eq.(\ref{vt}))  is not a sufficient
condition for  realizing a memory that is immune to  thermal
fluctuations.  A spectral equivalence exists between other TQO
inequivalent system pairs.  Our mappings also illustrate that the
quantum states themselves in a particular (operator language)
representation encode TQO and that the duality mappings, being non-local
in the original representation, disentangle the order. Thus, we cannot,
as coined by Kac, ``hear the shape of a  drum". The information is in
the eigenvectors.  

It was suggested \cite{entropy} that a deviation $\gamma>0$ from an
asymptotic area law scaling for the entanglement entropy is a measure
of TQO. We find that on its own, this (defined) measure of
``Topological Entanglement Entropy" -- a linear combination of
entanglement entropies  to extract $\gamma$ --  might not always be a
clear marker of TQO. Consider Klein spin models whose  ground state
basis is spanned by the product of singlet states \cite{Longversion}.
In most of these ground states, local measurements can lead to
different expectation values (and thus do  not satisfy Eq.
(\ref{def.})).  Still, we find \cite{Longversion} (i) an entanglement
entropy which deviates from an area law within the set of all ground
states that do not have TQO. Such an arbitrary -- contour-shape
dependent -- finite Topological Entanglement Entropy in a non-TQO
system is  clearly not in accord with the conjecture of 
\cite{entropy}. Conversely, (ii) a subset of TQO  ground states of the
Klein model constructed as linear superpositions of the singlet
product  states have zero entanglement entropy [$\gamma =0$].  The
criterion of \cite{entropy} deducing TQO from a finite $\gamma$ may
require additional improvements, such as the specification of an exact
limit and/or average for large contours. We conjecture that in order to
have TQO we need, at the very least, not only to have a finite 
Topological Entanglement Entropy but also that this entropy is
independent of the contour chosen.

The insufficiencies of the spectra and entropy for assessing if TQO is
present have counterparts in graph theory and in the Graph Equivalence
Problem in particular. The adjacency matrix of a graph has elements 
$C_{\bi\bj} =1$ if vertices $\bi$ and $\bj$ are linked by an edge and
$C_{\bi\bj}=0$ otherwise. Vertex relabeling $\bi \to p(\bi)$ leaves a
graph invariant but changes the adjacency matrix $C$ according to  $C
\to C^{\prime} =  P^{\dagger} C P$ with $P$  representing the
permutation: $P = \delta_{\bj,p(\bi)}$. The Graph Equivalence Problem
is the following \cite{godsil}: ``Given $C$ and $C^{\prime}$, can we
decide if both correspond to the same topological graph?"  The spectra
of $C$ and $C^{\prime}$ are insufficient criteria. Entropic measures
\cite{godsil} are useful but also do not suffice. 

In many systems (with or without TQO), there are non-local {\it string}
correlators which display enhanced (or maximal) correlations vis a vis
standard two-point correlation functions. We now outline an algorithm
for the construction of such non-local correlators. [In systems with
uniform global order already  present in their ground states, the
algorithm leads to the usual two-point correlators.] We seek a unitary
transformation $U_{s}$ which rotates the ground states into  a new set
of states which have greater correlations as measured by a  set of
local operators $\{V_{\bi}\}$. These new states may have  an
appropriately defined {\em polarization} (as manifest in the
eigenvalues $\{v_{\bi}\}$) of either (i) more slowly decaying
(algebraic or other) correlations, (ii) a uniform sign ({\em partial
polarization}) or (iii) maximal  expectation values $v_{\bi} = v_{max}$
for all $\bi$ ({\em maximal polarization}). Cases (i) or (ii) may lead
to a lower dimensional gauge-like structure for the enhanced
correlator. In systems with entangled ground states (such as in many
(yet not all) ground states with $T=0$ TQO),  $U_{s}$ cannot be a
uniform product of locals; $U_{s} \neq \prod_{\bi \in \Lambda} 
O_{\bi}$.  To provide a concise known example that clarifies these
concepts, we focus on case (ii) within the  well-studied  AKLT 
Hamiltonian \cite{AKLT}.  Here, there is a non-trivial unitary operator
$U_{s} \equiv \prod_{\bj<\bk} \exp[i \pi S^{z}_{\bj} S^{x}_{\bk}]$,
($[U_{s}, H] \neq 0$) which maps the ground states 
$\{|g_{\alpha}\rangle \}$  into linear superpositions of states in each
of which  the  local staggered magnetization 
$V_{\bi}=(-1)^{\bi}S_{\bi}^{z} \equiv  M_{\bi}$  is uniformly
non-negative (or non-positive) at every site $\bi$. As all transformed
states $|p_{\alpha} \rangle = U_s | g_{\alpha} \rangle$ are
superpositions of states with uniform sign  $v_{{\bf{i}}}$ (allowing
for two non-negative or non-positive  $v_{{\bf{i}}}$ values at every
site out of the three $s=1$ states),  $\tilde{G}_{\bi\bj}  \equiv 
\langle p_{\alpha} | V_{\bi} V_{\bj} |p_{\alpha} \rangle  > \langle
g_{\alpha}| V_{\bi} V_{\bj} |g_{\alpha} \rangle \equiv G_{\bi\bj}$.
When $U_{s}$ is written in full, $\tilde{G}_{\bi\bj}$ becomes  a
non-local {\it string} correlator $\tilde{G}_{\bi\bj} = 
(-1)^{|{\bf{i}}- {\bf{j}}|}  \langle g_{\alpha} | S_\bi^{z} \prod_{\bi
< \bk < \bj} \exp[i \pi S_{\bk}^{z}] S_{\bj}^{z} | g_{\alpha} \rangle =
(2/3)^{2}$ for arbitrarily large separation $|\bi-\bj|$. In the
$\otimes  S^{z}_{\bf{i}}$ eigen-basis, there are $(2^{N_{s}+1}-1)$
states $\ket{\phi}$  with non-negative (non-positive) $v_{\bi}$ for all
$\bi$ (they form a Hilbert subspace). Such an exponential number of
states appears in systems with a local $\Z_{2}$ gauge structure. Here,
the string correlators  exhibit a  symmetry under local $(d=0$) gauge
transformations while $H$ itself is not invariant.  These $d=0$
$\Z_{2}$ transformations correspond (in the $\otimes S^{z}_{\bi}$
basis) to the creation (annihilation) of an $S^{z}_{\bj}=0$ state at
any site ${\bj}$  followed by a unit displacement of $S^{z}_{\bf k}$ at
all sites ${\bf k}>{\bf j}$.   The operators linking the ground states 
form a  discrete, global, ($\Z_{2} \times \Z_{2}$) symmetry group
\cite{AKLT} and, as such, $T=0$ spontaneous symmetry breaking may
occur.  In the AKLT problem, the local expectation value $\langle
g_{\alpha} | S_{\bi=1}^{z} | g_{\alpha} \rangle$ depends on
$\ket{g_\alpha}$, violating Eq. (\ref{def.}) and suggesting that the
AKLT chain is not topologically ordered (nonetheless, in the open
chain there are spin-1/2 fractional excitations localized at the ends
of the chain). The general polarization operators are intimately tied
to selection rules which  hold for all of the ground states; these
selection rules (and a particular low-energy projection of truncated 
Hamiltonians for gapped systems) enable the construction of general
string and higher dimensional {\it brane}  correlators in general
gapped systems.  We may also generate maximal ($\tilde{G}_{\bi\bj} =1$)
string or brane  type correlators; here, the number of states with 
maximal polarization is finite and a non-local gauge-like structure
emerges. In systems with  known (or engineered) ground states, we may
construct polarizing transformations $U_{s}$. 

To conclude, we developed a symmetry-based framework to study physical
systems that display TQO. Symmetries may represent invariances of the
actual system Hamiltonian or of emergent low-energy phenomena.  In
particular, we proved a theorem establishing how $d \le 2$-dimensional
Gauge-Like Symmetries in gapped systems imply TQO (at all temperatures)
via freely propagating low-dimensional topological defects (e.g. $d=1$
solitons). This result extends the set of known systems which exhibit
TQO to include new orbital, spin, and Josephson  junction array
problems. All known TQO systems display Gauge-Like Symmetries
\cite{Longversion}. We examined physical consequences of those
symmetries (such as conservation laws, dimensional reduction, and
topological terms in high space dimension ($D$) theories), and 
low-dimensional gauge-like structures  in theories with entangled ground
states  in non-maximal string correlators (irrespective of TQO). We
discussed a general algorithm  for the construction of non-local {\it
string/brane} orders in systems where the ground states  conform to
certain selection rules and note that  {\it string/brane}  orders appear
in general gapped systems in arbitrary space dimensions.  We comment
that the Hamiltonian spectra, fractionalization,  entanglement entropy,
and maximal string correlations do not imply  TQO in all cases. We would
like to notice that a non-zero topological Chern number (the Hall
conductivity can be related to it \cite{thouless}) does not imply TQO
either, it only indicates that the system has a spectral gap with,
e.g.,  a  quantum state that breaks time-reversal symmetry
\cite{oursberry}, although the system can still host a Landau order
parameter. Similarly, many instances of degeneracy in TQO systems follow
from general time reversal invariance (Kramers' theorem); topological
considerations need not be invoked to establish ground state  degeneracy
-- especially so in the presence of a gap \cite{Longversion,nc}. In some
examples of TQO, an {\it exact dimensional reduction} occurs. For
example, the free energies of Kitaev's and Wen's $D=2$ models  are
equivalent to those of Ising chains, and consequently no
finite-temperature phase transition occurs \cite{Longversion}.  We also
drew analogies between TQO and the topology  of graphs. When unbroken by
external perturbations,  the $d$-dimensional Gauge-Like symmetries that
formed  the focus of our study may constrain the system dynamics.  Our
work extends and complements different studies of protected non-local
structures  elsewhere \cite{AharonovSchwartz}. 

We thank C. D. Batista, S. Bravyi, and P. Wiegmann for discussions, and
the DOE, NSA, CMI of WU for support.

\appendix

\section{Criteria for Robust topological quantum memory}
\label{app1}

Conditions for quantum error detection of TQO are reviewed in depth in
Ref. \cite{usnew}  where they are discussed for both zero and finite
temperatures. At $T=0$, they are given by
\begin{equation}
[\hat{P}_0 V \hat{P}_0, \hat{T}_\mu]=0 ,
\label{detection_cond}
\end{equation} 
where $\hat{P}_0=\sum_\alpha \proj{g_\alpha}$ is the protected
subspace, and $\hat{T}_\mu$'s represent the logical operators.  These
operators are ($d$-dimensional Gauge-Like) symmetries of $H$, i.e.  
\begin{eqnarray}
[H,\hat{T}_{\mu}]=0 ,
\label{good}
\end{eqnarray} 

The propagation of a local error ($ V $) at finite temperature causes no
harm to the non-local logical operators $\{\hat{T}_{\mu}\}$ and the
algebra that they satisfy. The condition of Eq. (\ref{detection_cond})
can be related to the TQO condition of  Eqs. (\ref{def.}) \cite{usnew}.
Any system with a unique ground state must automatically satisfy
$[\hat{P}_{0}V\hat{P}_{0},\hat{T}_{\mu}] = 0$. This is so as
$\hat{P}_{0}V \hat{P}_{0}  = v \hat{P}_{0}$ with $v$ a $c$-number: $v=
\langle g|V|g \rangle$ where $|g \rangle$ is the unique ground state.
Thus, $[\hat{P}_{0}V \hat{P}_{0},\hat{T}_{\mu}] = v[ \hat{P}_{0},
\hat{T}_{\mu}]$. As $\hat{P}_{0} = f(H)$ (a function of the Hamiltonian)
and as Eq. (\ref{good}) holds, we have that $[\hat{P}_{0},\hat{T}_{\mu}]
=0$. This shows that Eq. (\ref{detection_cond}) always holds in the
absence of degeneracy. In all topological quantum computation
applications, the symmetries $\{\hat{T}_{\mu}\}$ do not all commute with
one another. The system must then exhibit  degeneracy. In those
applications, the symmetry  operators $\hat{T}_{\mu}$  represent
non-commuting braiding operations.

\section{Symmetry-based analysis of autocorrelation times and crossover temperatures}
Below, we briefly elaborate on autocorrelation times and cross-over temperatures
in general systems with $d$ dimensional gauge like
symmetries \cite{usnew, A2}.
 In general, we may apply the bounds of \cite{usnew, BN, Longversion}
to arrive at both upper and lower bounds on all correlation lengths
and auto-correlation times in general systems with
$d$ dimensional gauge like symmetries. Kitaev's toric code model 
has a $d=1$ Ising symmetry. As we showed \cite{A2, Longversion}, Kitaev's model is
special in that it may be exactly 
mapped to two decoupled Ising chains (a system of dimension 
$d=1$). This mapping realizes a particular case in which the bounds of \cite{BN} become exact
equalities.  Our general considerations are, however, not limited to such special solvable models  
and follow from the symmetry related dimensional reduction as  it is manifest
in the bounds of \cite{BN}.  We showed \cite{usnew} that the finite temperature equilibrium expectation
values of all quantum memory based stabilizer operators vanish in any spatial dimension.

For illustrative purposes, we discuss systems
such as Kitaev's model that have a discrete 
$d=1$ Ising symmetry. By the bounds
of \cite{BN} and in this particular case, the 
exact mapping of \cite{A2, Longversion} enables us to relate
the problem to its counterpart in a $(d=1$) Ising 
chain. We focus on quantities such as the Kitaev model toric
code cycle operators that are not invariant under all $d=1$ Ising symmetries
and may consequently be destroyed by $d=1$ Ising domain walls.
On a single Ising chain with $N_{s}$ spins, the correlation length
gives rise to a cross-over temperature 
scaling as $1/\ln N_{s}$ \cite{A2}.
Such a crossover temperature may be equivalently derived
from energy-entropy balance considerations in the free energy,
$F = E - TS$, for the insertion of a single $d=1$ defect (i.e., a domain
wall for discrete symmetries) in a system with short range interactions.  
Associated with the insertion of  a single defect there is 
an energy penalty $E = \Delta$, and an entropy 
$S \sim \ln N_{s}$. These two contributions lead to a
a crossover temperature scaling as $T_{\sf cross} \sim 1/\ln N_{s}$, 
i.e., for $T<T_{\sf cross}$ defects are unfavorable. 
In accord with the one-dimensional character, the 
low temperature autocorrelation times may  scale as
$\tau \sim \exp[\Delta/(k_{B} T)]$.  By an extension of the bounds of \cite{BN, Longversion}, the autocorrelation
function in a general system with $d$ dimensional gauge like symmetries,
is bounded (both from above and from below) by autocorrelation
functions of a $d$ dimensional system in which the range of
the interactions and symmetries are preserved.
It is notable, as discussed in the text, that in systems with interactions of finite range
and strength the autocorrelation
times of quantities not invariant under  $d=1$ gauge-like 
symmetries do not scale with the system size \cite{Longversion}.
Thus, in such systems (including, but not limited
to, Kitaev's and Wen's models), the autocorrelation
times are system size independent. In a similar
fashion,  system size dependent 
auto-correlation times in the ordered phase appear in low ($d$)
dimensional systems with long range 
interactions that exhibit a phase transition such as ferromagnetic
spin chains with $1/r^{\lambda}$ interactions ($\lambda \le 2$)
\cite{AYH}. Taken together with the bounds of \cite{usnew,BN,Longversion}, this allows for
divergent autocorrelation times of quantities that are not invariant under $d$ dimensional
gauge like symmetry in systems
that have long range interactions.

\section{Gauge Like Symmetries of the Quantum Dimer Model}
In \cite{Longversion}, we analyzed in some depth the gauge like symmetries (exact or emergent) in many systems. 
Below we further briefly discuss exact gauge like symmetries of the quantum dimer model \cite{QDM}, as it is manifest in a
gauge theory formulation \cite{NN}.  This discussion elucidates the known
degeneracy between ``even" and ``odd" sectors of this model and shows
how to derive the degeneracy (for the entire spectrum of the theory) by formally formulating the 
exact gauge like symmetry of the quantum dimer model.  
We will follow the notation of \cite{NN} with
$\sigma^{z}_{ij} = 1$ denoting
the presence of a dimer on the link connecting sites $i$ and $j$ and $\sigma^{z}_{ij}  =-1$ its absence.
It is readily seen that the quantum dimer model Hamiltonian of Eqs. (4,5) in \cite{NN} is invariant under
exact symmetries- both local ($d=0$) symmetries of the form $A_{i} = \prod_{j} \sigma^{z}_{ij}$
with $j$ denoting all nearest neighbors of $i$ as well as 
the ($d=1$ dimensional) toric cycle operators $Z_{1,2} = \prod_{ij \in C_{1,2}} \sigma^{z}_{ij}$.
The latter symmetries involve (i) for $Z_{1}$ the product of $\sigma^{z}$ on all horizontal bonds on a column that goes around one 
cycle of the torus (the contour $C_{1}$) or  (ii) for $Z_{2}$ the product  of all $\sigma^{z}$ on all vertical
bonds along a row that goes around the other cycle of the torus (the contour $C_{2}$).
For a square lattices of size $L_{x} \times L_{y}$ on a torus with both $L_{x}$ and $L_{y}$ odd,
the time reversal symmetry inverts the sign of the toric cycle symmetries $Z_{1,2} \to -Z_{1,2}$.
These exact cyclic symmetries flesh out the even and odd sectors (and their degeneracy) of the entire spectrum
of the theory.


\begin{thebibliography}{99}
\bibitem{timereversal}
The $d=1$ symmetry operators $O^{\alpha}$ are the minimal symmetry
operators that link ground states. A global symmetry operation such as
time reversal is a composite of these more basic symmetry operators,
i.e., it does not link more states. The same applies to other global
symmetries. Let us illustrate this property:   The time reversal
operator is given by
\begin{eqnarray}
\Theta = \eta \prod_{r}(-i \sigma^y_r) K = \eta' \Big( \prod_{C_{x}}
O^x_{C_{x}} \Big) \Big( \prod_{C_{z}} O^z_{C_{z}}  \Big) K,
\label{czx}
\end{eqnarray}
with $\eta \ (\eta')$ being phase factor $c$-numbers, and $K$ complex
conjugation. The products in Eq. (\ref{czx}) are over all rows
($\{C_{x}\}$) and columns ($\{C_{z}\}$). Each lattice site $r$ appears
in one vertical column and one horizontal row. As the orbital compass
model Hamiltonian is real (in the basis of $\sigma^{z}$), $[H,K]=0$; as
a result $\Theta$ commutes with $H$. The ground state degeneracy is a
consequence of the fact that $[O^x_{C_{x}},O^z_{C_{z}}]\neq 0$. Kramers
degeneracy may be realized in systems with or without TQO.  In other
words, time reversal is a consequence of the $d=1$ symmetries
$O^{\alpha}$ (and the reality of the Hamiltonian), and  it does not lead
to additional relations not present in the $d=1$ symmetries. The theorem
and all results that are to be presented in this work are adduced from
the minimal set of symmetry operators (i.e., symmetry operators that are
smallest in their spatial (local) support) that suffice to link all
ground states.
\end{thebibliography}
\end{document}